\documentclass{article}

\usepackage{arxiv}

\usepackage[utf8]{inputenc} 
\usepackage[T1]{fontenc}    
\usepackage{hyperref}       
\usepackage{url}            
\usepackage{booktabs}       
\usepackage{amsfonts}       
\usepackage{nicefrac}       
\usepackage{microtype}      
\usepackage{lipsum}
\usepackage{graphicx}
\usepackage{subfigure}
\usepackage{amssymb}
\usepackage{amsmath}
\usepackage{bm}
\usepackage{multirow}
\usepackage[OT1]{fontenc} 
\usepackage{algorithm,algpseudocode}
\usepackage{amsmath}
\usepackage{url}
\usepackage{color}

\title{Referenceless Rate-Distortion Modeling with Learning \\ from Bitstream and Pixel Features}

\author{
  Yangfan Sun \\
  University of Missouri-Kansas City \\
  Kansas City, MO 64110 \\
  \texttt{ysb5b@umsystem.edu} \\
   \And
    Li Li \\
  University of Missouri-Kansas City \\
  Kansas City, MO 64110 \\
  \texttt{lil1@umkc.edu} \\
  \And
  Zhu Li \\
  University of Missouri-Kansas City \\
  Kansas City, MO 64110 \\
  \texttt{lizhu@umkc.edu} \\ 
  \And
  Shan Liu \\
  Tencent America \\
  Palo Alto, CA, 94306 \\
  \texttt{shanl@tencent.com}\\ 
  }

\begin{document}
\maketitle

\begin{abstract}
Generally, adaptive bitrates for variable Internet bandwidths can be obtained through multi-pass coding.
Referenceless prediction-based methods show practical benefits compared with multi-pass coding to avoid excessive computational resource consumption, especially in low-latency circumstances.
However, most of them fail to predict precisely due to the complex inner structure of modern codecs.
Therefore, to improve the fidelity of prediction, we propose a referenceless prediction-based R-QP modeling (PmR-QP) method to estimate bitrate by leveraging a deep learning algorithm with only one-pass coding.
It refines the global rate-control paradigm in modern codecs on flexibility and applicability with few adjustments as possible.
By exploring the potentials of bitstream and pixel features from the prerequisite of one-pass coding, it can reach the expectation of bitrate estimation in terms of precision.
To be more specific, we first describe the R-QP relationship curve as a robust quadratic R-QP modeling function derived from the Cauchy-based distribution. 
Second, we simplify the modeling function by fastening one operational point of the relationship curve received from the coding process.
Third, we learn the model parameters from bitstream and pixel features, named them hybrid referenceless features, comprising texture information, hierarchical coding structure, and selected modes in intra-prediction.
Extensive experiments demonstrate the proposed method significantly decreases the proportion of samples' bitrate estimation error within $10\%$ by $24.60\%$ on average over the state-of-the-art.
\end{abstract}

\keywords{Rate-distortion modeling; referenceless rate-distortion model; machine learning; transcoding; video processing}





\maketitle

\section{Introduction}
\label{Sec::introduction}
The knowledge of bitrate and corresponding video quality is the necessary prerequisite to make the optimal bitrate allocation strategies for Internet-based video services, otherwise, it may cause a large of unnecessary waste or deficiency on clients' bandwidths. 
However, due to the inner complexity of modern codecs, e.g., high efficient video coding (HEVC)~\cite{sullivan2012overview} and advanced video coding (AVC)~\cite{wiegand2003overview}, it has become a challenge to assess the bitrate and video quality in an accurate and fast way.

Many research focused on the characteristics of bitrate and quantization parameter (R-QP) of block-level rate-control paradigm~\cite{chiang1997new}~\cite{wang2013quadratic}, which require a lot of inner algorithm adjustments for different codecs to execute bit allocation. 
It might cause fluctuation of temporal qualities due to insufficient bit assignment on few last coding units (CUs) of coding frames. 
To solve this issue, we adopt a global-based rate-control paradigm that considers each video clip (or frame) as a basic unit~\cite{covell2016optimizing}~\cite{sun2018machine}~\cite{sun2020Yoco}. 
It can work with most modern codecs without the need for excessive block-level adjustments, that provide a global strategy of bit assignment to prevent inconsistent video quality. 
Moreover, parallel implementation can be achieved on video clips (or frames) in proportion to available computational resources. 

The global rate-control paradigm certainly has many inherent advantages over the block-level paradigm.
However, previous attempts have shown the fundamental problem of describing the characteristics of factors in bitrate allocation of the global paradigm.
This bottleneck is mainly caused by the following reasons:
1) The global multi-pass coding method can establish the actual R-QP relationship curve, but excessive computational cost is needed.
2) Insufficient content or coding information was adopted to describe the R-QP relationship, e.g., Xu et al.~\cite{xu2017cnn} and Santamaria et al.~\cite{santamaria2018estimation} only used pixel information (original frames).
Covell et al.~\cite{covell2016optimizing} and Sun et al.~\cite{sun2018machine} simply took statistic coding domain data into account.
3) The linear modeling function is insufficient to fit nonlinear R-QP relationship.
While considering situations of multiple resolutions or frame sizes, the fitting performance of the linear model~\cite{covell2016optimizing} deteriorates as the resolution of frame size increases~\cite{sun2018machine}.

\begin{figure*}[tb]
\centering
\includegraphics[width=1\textwidth]{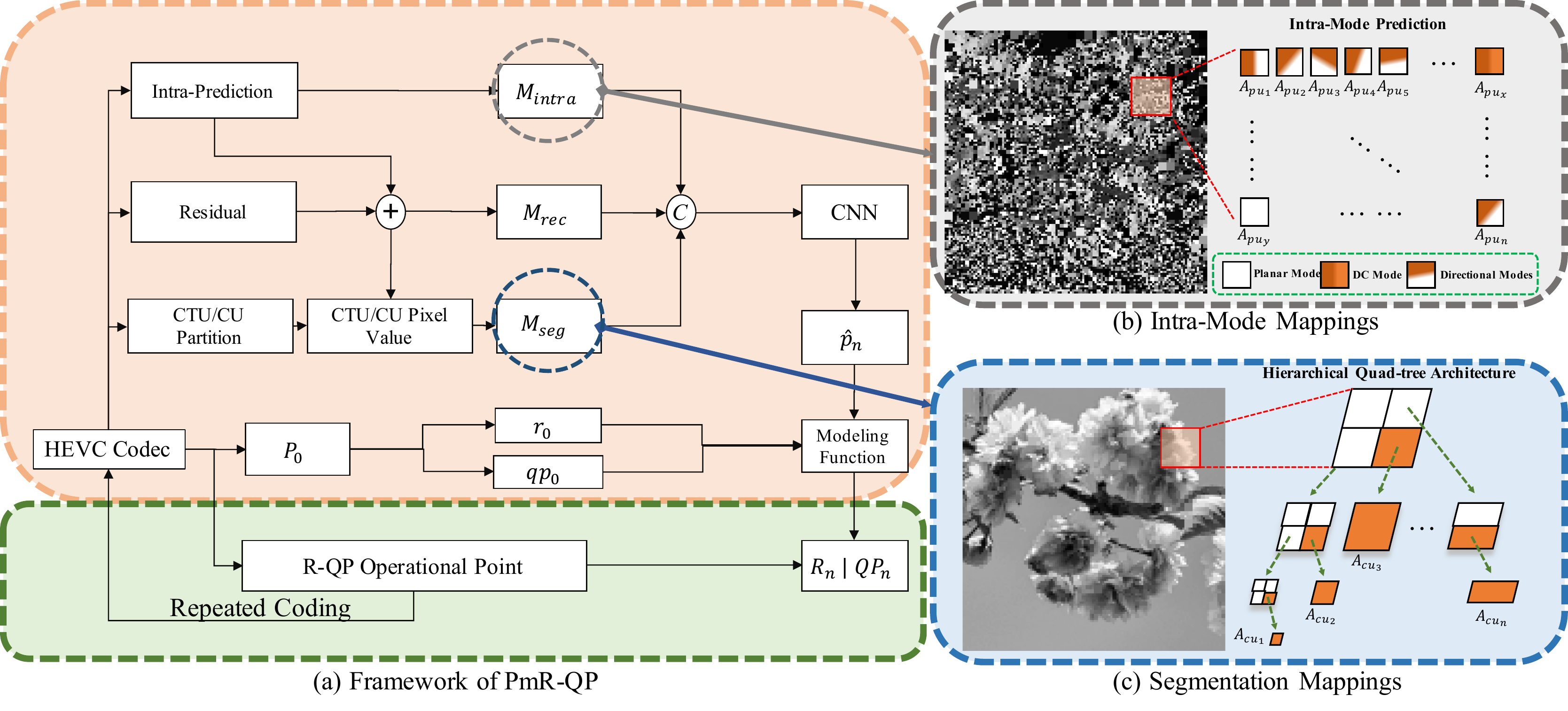}
\caption{(a) Framework of PmR-QP compared with multi-pass coding method. (b) Extraction procedure of intra-mode mappings $M_{intra}$. (c) Extraction procedure of segmentation mappings $M_{seg}$.}
\label{Fig::Framework}
\end{figure*}
Therefore, to solve this issue, we propose a referenceless prediction-based R-QP modeling method (PmR-QP) throughout only one-pass coding needed as a prerequisite to extract the bitstream and pixel information.
It exploits the potentials of hybrid referenceless features to track down the quantization processes, which are used to signal redundant information elimination. 
Then, leveraging a deep learning algorithm as the replacement of actual coding~\cite{huang2018qarc}~\cite{huang2019generalizing}, PmR-QP learns the corresponding oblique relationships between the extracted representatives and bitrate-quality information.
To be specific, firstly, we develop an optimized R-QP modeling function to characterize the relationship between bitrate and QP.
Secondly, we enhance and unify the features of bitstream in multiple coding domains to learn the content-dependent R-QP model parameters from scratch.
Currently, we successfully validate the efficiency of PmR-QP in intra-predicted frames, which occupy the majority proportion of bits in videos.
The contributions of this paper list as followed,
\begin{itemize}
\item{
We derive a quadratic R-QP modeling function from the Cauchy-based distribution to characterize the relationship between bitrate and corresponding QP, which is used to directly control video quality.
The quadratic modeling function can fit the non-linear R-QP relationship better than the previous linear rate-control modeling function~\cite{covell2016optimizing}.
We have it tested to prove its feasibility in real cases prevailingly.
}
\item{
We fasten an operational point on the R-QP relationship curve from the one-pass coding in passing that no additional computational cost is needed.
As the means of model parameter elimination, it can greatly improve the proceeding of deep learning in inferring speed and estimated accuracy.
}
\item{
We significantly explore the potentials of bitstream and pixel information from multi-levels coding domains, e.g., reconstruction, hierarchical segmentation, and macro-block intra-prediction.
To concatenate them in the proposed network, we modify these features to a uniform type and structure.
To the best of our knowledge, no previous works have used the homogeneous scheme because of inconsistency in different coding domains.
}
\item{
Performance experiments and ablation studies on DIV2K dataset \cite{agustsson2017ntire}, demonstrate the PmR-QP method outperforms the state-of-the-art linear modeling solution. 
}
\end{itemize}

The remainder of the paper is organized as follows.
Section~\ref{Sec::Related Work} will introduce related researches.
In Section~\ref{Sec::proposed method}, we will discuss the proposed R-QP modeling function, and hybrid bitstream features in details.
The proposed network and hyper-parameters will be discussed in Section~\ref{Sec::Architecture of Proposed Network}.
Section~\ref{Sec::Experimental Results} will show the detailed experimental results and Section~\ref{Sec::Conclusion} will present the conclusions and future plans. 

\section{Related Work}
\label{Sec::Related Work}
The knowledge of rate and distortion (R-D) relationship is essential for rate control.
It decides how many bits should be provided to obtain minimal distortion subject to the budget of bits.
Ou et al.~\cite{ou2011ssim} considered a similarity index as a quality metric for R-D model to correlate bitrate allocation with human perception.
Gao et al.~\cite{gao2016ssim} proposed a Nash bargaining solution for optimizing a structural similarity index (SSIM)-based CTU-level R-D scheme.
Both of these methods need the actual R-D relationship from multiple passes of coding.
Because the R-D data is recorded from real samples, the accuracy can be assured.
However, the excessive computational cost needs to be reduced while being applied to latency-sensitive scenarios.
Therefore, differing from actual coding, the idea of R-D estimation was proposed for more practical video applications.

Estimation methods can be divided into two categories depending on whether adopting a modeling function to calibrate the estimated results or not.
Non-modeling-based methods implement end-to-end frameworks to predict R-D relationship directly, while modeling-based methods derive R-D relationship from modeling functions.
Many researchers have leveraged deep learning algorithms to estimate R-D relationship due to their availability in different video applications~\cite{laude2016deep}~\cite{yang2018using}~\cite{jia2017spatial}~\cite{yang2007advances}.
For non-modeling-based methods, Xu et al.~\cite{xu2017cnn} and Santamaria et al.~\cite{santamaria2018estimation} proposed CNN-based R-D estimated methods. 
They both adopted original frames as references to estimate the R-D relationship explicitly.
In~\cite{xu2017cnn}, SSIM maps were attributed as distortion and learned through a novel CNN in separate with the number of bits.
Santamaria et al.~\cite{santamaria2018estimation} followed a similar framework to estimate the number of bits (pixel-wise) and absolute distortion mappings instead of SSIM maps individually, through a neural network with two pipelines.
The notion of nonlinearity for R-D estimation was proposed in~\cite{santamaria2018estimation} and activated function was improved by adding Parametric Rectified Linear Unit (PReLU)~\cite{he2015delving} to achieve nonlinear fitting.
These solutions diminished the complexity of codex over multi-passes coding.

Then, modeling-based R-D estimation methods were proposed by Covell et al.~\cite{covell2016optimizing} and Sun et al.~\cite{sun2018machine}.
Covell et al.~\cite{covell2016optimizing} used statistic coding representatives to predict bitrate and constant rate factor (CRF) implicitly through a linear logarithmic R-CRF model.
Sun et al.~\cite{sun2018machine} optimized the R-CRF model to second-order function, which described the nonlinearity of R-CRF relationship better.
Both of them only employed pure statistical coding information, which has been proved its insufficiency to describe video content.
On the other hand, the disadvantage of current non-modeling-based methods was the mere adoption of pixel information (frames).
Neither of them used these intersectional domains data to study the R-D relationship. 
The boundedness of single domain data might mislead the algorithms to make a global decision.

\section{Proposed Method}
\label{Sec::proposed method}
\subsection{Overall Framework}
\label{subsec::Overall Framework}
In this section, we elaborate upon the framework of PmR-QP method.
As aforementioned, QP is adopted as the quality metric of intra-frames and used to directly control bitrate by the employed codec.
Fig.~\ref{Fig::Framework} shows the details of the framework, whose objective is to predict the content-dependent R-QP model parameters $\hat{p_n}$ initially, then derive the corresponding bitrate through the proposed model with the given QP.
As shown, the complete framework can be divided into three subtasks, R-QP modeling function $m(.)$, referenceless features extraction $E_b(.)$, and network training $t(.)$.
With the given $QP$ and trained R-QP model parameters $\hat{p_n}$, we can derive the predicted bitrate $\hat{R}$ as followed,
\begin{equation}
\label{Eq::rate control modeling formulation}
\hat{R} = m(QP,\hat{p_n}).
\end{equation}

$\hat{p_n}$ are learned from the concatenated hybrid referenceless features,
\begin{equation}
\label{Eq::network}
\hat{p_n} = t(cat(M_{rec},M_{seg},M_{intra}), \Theta),
\end{equation}
where $\Theta$ denotes the set of network variables (weights and biases).
$M_{rec}$, $M_{seg}$, and $M_{intra}$ represent the hybrid referenceless features (reconstructed, segmentation, and intra-mode mappings).
They are extracted from the target intra-frames in multiple coding domains,
\begin{equation}
\label{Eq::bitstream patterns extraction}
[M_{rec},M_{seg},M_{intra}] = E_b(\gamma),
\end{equation}
In summary, Eq.~(\ref{Eq::rate control modeling formulation}), Eq.~(\ref{Eq::network}), and Eq.~(\ref{Eq::bitstream patterns extraction}) can jointly merge to the PmR-QP method, denoted as $F_{p}(.)$,
\begin{equation}
\label{Eq::overall}
\hat{R} = F_{p}(\gamma).
\end{equation}

\subsection{Proposed R-QP Modeling Function}
\label{subsec::Proposed R-QP Modeling Function}
The existing study in~\cite{kamaci2005frame} clarified the knowledge of Discrete cosine transform (DCT)'s coefficients' probability distribution is critical at the derivation of the relationship between bitrate and Quantization step (Qstep) (associated with QP).
Here, to explore an accurate description of the bitrate and QP relationship, we formulate the R-QP modeling function derived from the entropy of DCT's coefficients.

Generally, bit allocation strategy splits bits into two groups: header bits $R_{h}$ and residual bits $R_{r}$ (dominant fraction of total bit consumption $R_{total}$).
We can assume that
\begin{equation}
\label{Eq::rate_total_to_rate_residual}
R_{total} \approx R_{r}.
\end{equation}

As known, $R_{r}$ is related to the entropy of DCT's coefficients.
Due to the property of residual bits, the entropy of DCT's coefficients is extremely sensitive to quantization.
Therefore, the approximate correlation of total bits $R_{total}$ and the entropy of DCT's coefficients $H(Q)$ at varying $Q$ (Qstep) can be represented as,
\begin{equation}
\label{Eq::eentropy_to_rate}
H(Q) \approx R_{total}(Q) \approx R_{r}(Q).
\end{equation}

In~\cite{kamaci2005frame}, the probability distribution function (PDF) of Cauchy distribution \cite{eggerton1986statistical} is proven that a better description is on actual data than Gaussian \cite{goodman1963statistical} and Laplacian \cite{muller1993distribution} distributions.
Then, the entropy of quantized DCT's coefficients in informative theory can be extended based on Cauchy-PDF,
\begin{equation}
\label{Eq::entropy_cauchy_derivation}
\begin{split}
H(Q) = - \frac{1}{\pi} \sum_{n=-\infty}^\infty  tan^{-1} \frac{\gamma Q}{\gamma^2 + (n^2 -0.25)Q^2}\\
\times log_2{[\frac{1}{\pi} tan^{-1} \frac{\gamma Q}{\gamma^2 + (n^2 -0.25)Q^2}]}, \\
n = \pm{1}, \pm{2} ..., \pm{N},
\end{split}
\end{equation}
where, $nQ$ denotes as quantization level, while $\gamma$ is the variable of zero-mean Cauchy-PDF.
\begin{figure}[tb]
\centering
\includegraphics[width=0.4\textwidth]{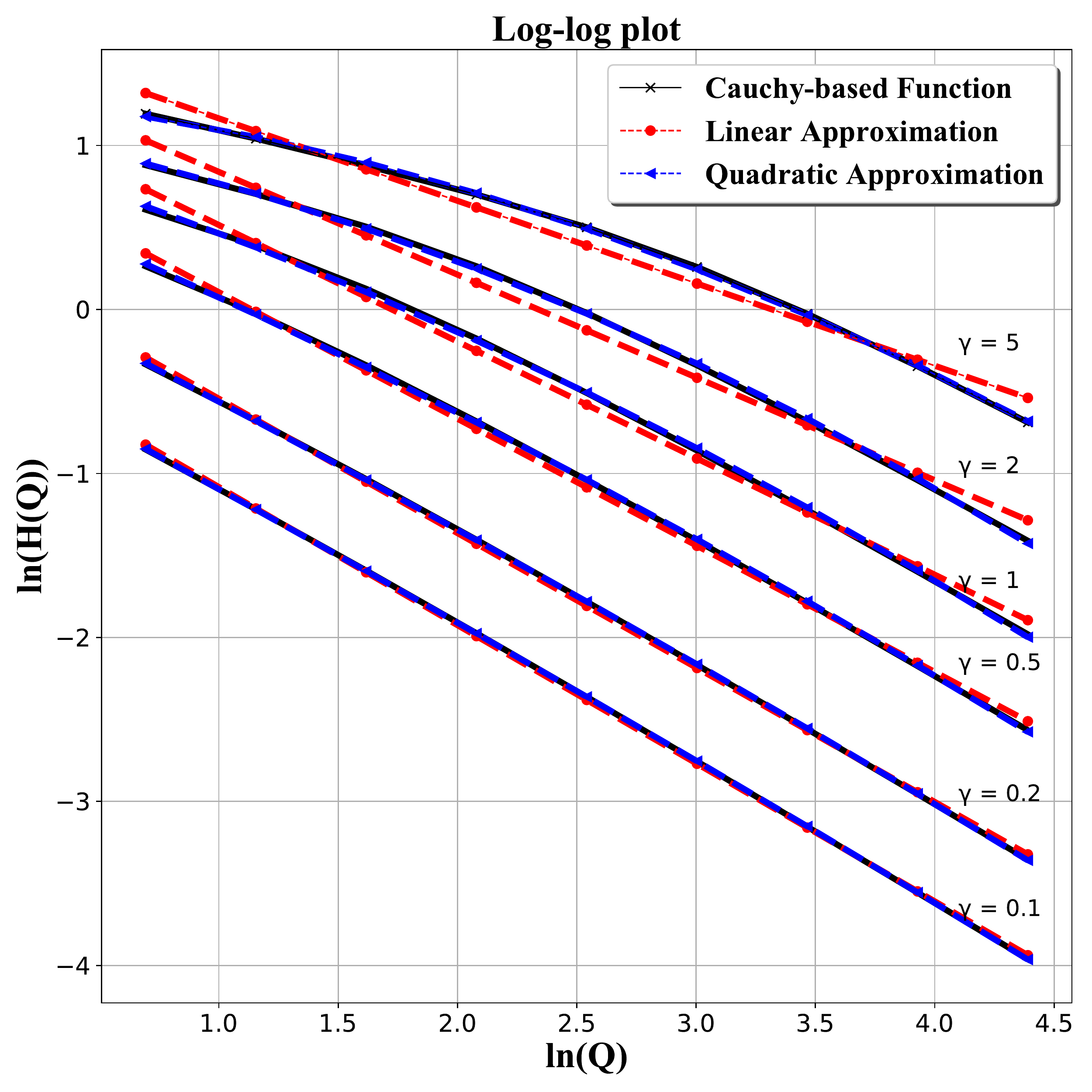}
\caption{H(Q)-Q log-log plot of Cauchy-PDF-based function approximation.}
\label{Fig::cauchy_based}
\end{figure}
A linear modeling function between H(Q) and $Q$ \cite{ushakov2001density} was proposed to simplify Eq.~(\ref{Eq::entropy_cauchy_derivation}),
but hardly to characterize the non-linear $H(Q)-Q$ relationship, especially when $\gamma$ increases at a large margin.
The hypothesis quadratic $H(Q)-Q$ relationship can be suggested given by the observation of Fig.~\ref{Fig::cauchy_based},
\begin{equation}
\label{Eq::q_approx}
ln(Q) \propto ln(H(Q))^2 + ln(H(Q)) + c, 
\end{equation}
where $c$ represents a constant.
The transformation between $QP$ and $Q$ can follow the equation below,
\begin{equation}
\label{Eq::qstep_to_qp}
QP = 6 \cdot log_2{Q} + 4, 
\end{equation}
Since we need to assess the relationship between bitrate and $QP$ eventually, Eq.~(\ref{Eq::eentropy_to_rate}), Eq.~(\ref{Eq::q_approx}) and Eq.~(\ref{Eq::qstep_to_qp}) can be joint derived that the entropy of DCT's coefficients in varying $QP$ has a quadratic logarithmic changing trend in approximation,
\begin{equation}
\label{Eq::r_to_qp}
QP \propto \frac {6 [ln(R(QP))^2+ ln(R(QP))]-4}{ln(2)}. 
\end{equation}

Therefore, based on previous R-D modeling functions \cite{he2001low} \cite{li2014} \cite{li2013qp} \cite{covell2016optimizing}, we devise a quadratic logarithmic modeling function from Eq.~(\ref{Eq::r_to_qp}),
\begin{equation}
\label{Eq::second_r_qp}
QP = \alpha(\gamma)ln(R(QP))^2 + \beta(\gamma)ln(R(QP)) +\mu(\gamma),
\end{equation}
where $\alpha(\gamma)$, $\beta(\gamma)$, and $\mu(\gamma)$ denote as content-dependent model parameters related to intra-prediction frame $\gamma$.

The proposed quadratic R-QP modeling function can fit a wide range of QP settings to achieve many practical uses precisely.
But the increasing number of model parameters results in prohibitive levels of training complexity compared with the linear approximation~\cite{covell2016optimizing}.
To overcome this issue, we further explore the potential of coding information to simplify the modeling function.
An operational R-QP point $P_0$ is encoded along with bitstream data that have been ignored previously.
We decide to involve it by fastening the proposed modeling function on $P_0$.
$P_0$ would not deteriorate the fitting capacity of function since it is from actual coding.
Meanwhile, the freedom of fastened function is limited fractionally with fewer model parameters to learn.
Assume the values of bitrate and QP at $P_0$ are $r_0$ and $qp_0$, respectively, then R-QP modeling function can be eliminated one modeling parameter as followed,
\begin{equation}
\label{Eq::final_r_qp}
\begin{split}
QP = \alpha^{*}(\gamma)[ln(R(QP))^2-ln(r_0)^2] +  \\
 \beta^{*}(\gamma)[ln(R(QP))-ln(r_0)] + qp_0,
\end{split}
\end{equation}
where $\alpha^{*}(\gamma)$ and $\beta^{*}(\gamma)$ are the model parameters pending to predict from network after simplification.
This proposed modeling function successfully balances the trade-off of training difficulty and predicted precision, evaluated in Sec. ~\ref{subsubsec::Improvements in Modeling Function Simplification}.
\begin{figure*}[tb]
\centering
\includegraphics[width=1\textwidth]{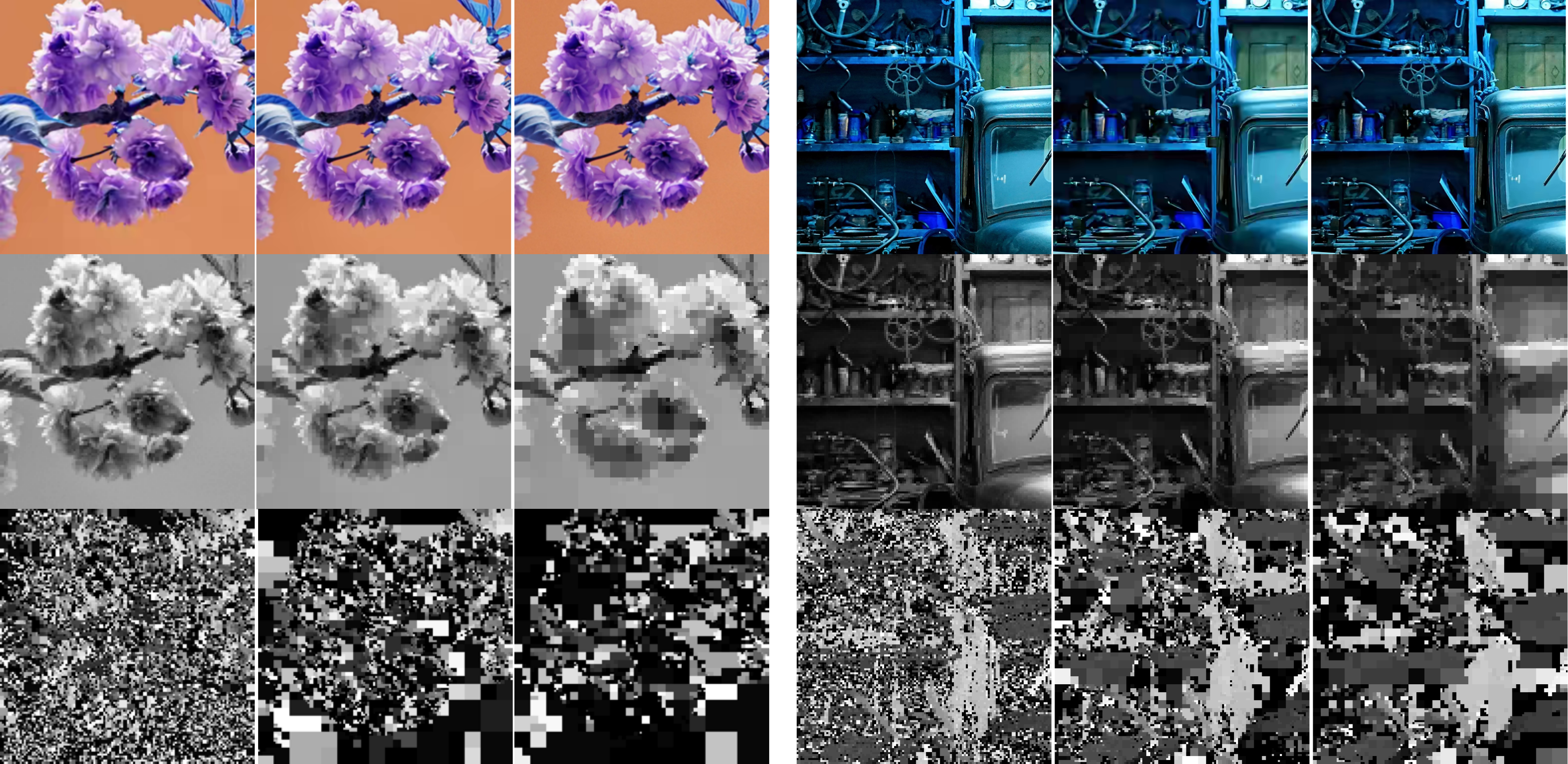}
\caption{Hybrid bitstream features in QP settings $\in$ \{10, 26, 38\} in different levels of texture complexity, including $M_{reg}$, $M_{seg}$, and $M_{intra}$, respectively.}
\label{Fig::bitstream_patterns}
\end{figure*}

\subsection{Hybrid Referenceless Features}
\label{subsec::Hybrid Bitstream Representatives}
The core execution of compression techniques in most modern codecs is to eliminate overlapped or redundant information in terms of spatial, temporal, statistical, and visual domains~\cite{lei2000video}.
Bits can be saved due to the eliminating processes without abandoning any relevant data.
Therefore, as long as we detect the exact number of bits saved subject to corresponding levels of quality in compression, the relevance of bitrate and QP can be received in different levels of quantization.
However, due to the complex inner architectures of modern coding standards, such as HEVC, it is extremely difficult to approximate the saved bits~\cite{bossen2012hevc}. 
Due to the successes of deep learning algorithms in many multimedia applications, it is natural to come up with a learning-based method as the fundamental core to track down the saved bits by exploring the features of information redundancies in different coding procedures.
However, the type and structure of data in different coding procedures is inconsistent to analyze, which leads that most previous researches only concentrated on the studies of local procedures partially.

In this paper, we bring out the unification of the type and structure of data in different coding domains.
It takes advantage of the global features extraction of bitstream and pixel information, as hybrid referenceless features, to significantly improve the estimated performance.
In detail, the hybrid referenceless features include the components of texture information, hierarchical coding structure information, and intra-predicted modes.
We unify them into pictures by mapping them to two-dimension planes, as shown in Fig.~\ref{Fig::bitstream_patterns}.

Fig.~\ref{Fig::bitstream_patterns} demonstrates hybrid referenceless features from two intra-predicted frame samples with different levels of texture intricacy.
We visualize features by coding in different QP settings (QP=$\{$10, 26, 38$\}$) to study the quantized sensitiveness of each feature individually.
Initial assumption indicates the referenceless features mirror the progression of quantization that leads the possibility to learn the property of R-QP through the exploration of referenceless features.
The observation shows that underlying layers (segmentation and intra-prediction) of coding respond to the changes of QP more unmistakably.
Even though their descriptions of the multifaceted nature of images and the distribution of high and low frequencies are not as good as the reconstructed image, they can increase the quantized sensitiveness of the referenceless features.
The following paragraphs will explain the generalization of each feature step by step.

\subsubsection{\textbf{Reconstructed Mappings}}
Reconstructed Mappings $M_{rec}$ possess the homologous structure with corresponding original images since the quantization would not damage the integrity of reconstructed Mappings $M_{rec}$ but simply modify the number of bits for each symbol~\cite{mccarthy2004sharp}.
They are defined as the representatives in pixel-domain to preserve the texture information in low-distorted details.
As the replacement of original images, reconstructed mappings $M_{rec}$ are extracted to describe the content complexity and distribution of high- and low-frequency information.
As shown in Section~\ref{Sec::Experimental Results}, they dominate the circumstances of a single feature as input.
It proves the efficiency of $M_{rec}$, especially lacking original data, e.g., video transcoding \cite{xin2005digital}.

\subsubsection{\textbf{Segmentation Mapping}}
The quad-tree coding tree units (CTUs) architecture is employed with variable sizes of units~\cite{ohm2012comparison} in HEVC codec, which can be partitioned into hierarchical coding units (CUs) and further divided into predicted units (PUs).
The availability of larger block sizes in a quad-tree partitioning structure decides the most significant improvement of coding efficiency compared with previous codec generation~\cite{kim2012block}, also preserving dominant bits saving.
The knowledge of CU partition might help track the arrangement of bits at macroblock (MB) level~\cite{purnachand2012fast}.
For instance, a larger-size CU requires less bit per pixel (bpp) than a smaller-size CU, while a deeper depth of CU requires more bpp.

To explore the partitioning information, we extract and visualize CU information, denoted them as segmentation mappings $M_{seg}$, as shown in Fig.~\ref{Fig::Framework}(c). 
To be specific, three types of partitioning factors are utilized, including the average pixel esteem, size, and location of each CU.
We first create a sharing-sized blank image with the original frame and split it into number of different scaled areas according to the corresponding size and location of CUs.
The average pixel esteem is assigned as the shared pixel values for the complete scaled areas, as shown
\begin{equation}
\label{Eq::segmentation}
M_{seg} = \O_{r}{([A_{cu_{1}}]_{w_1 \times h_1}, [A_{cu_{2}}]_{w_2 \times h_2}, ..., [A_{cu_{n}}]_{w_n \times h_n})},
\end{equation}
where,
\begin{equation}
\label{eq::linear model}
  A_{cu_{n}} = \sum^{w_n}_{i} \sum^{h_n}_{j} \frac{V_{p(i,j)}}{w_n h_n}.
\end{equation}
Here, the sets of $\{w_1, w_2...w_n\}$ and $\{h_1, h_2...h_n\}$ represent the width and height of corresponding CUs, respectively.
${\O_r}(.)$ is the reshaping operator and $V_{p(i,j)}$ denotes as the pixel value at location of $\{i,j\}$ in $M_{rec}$.
The operator $[.]_{w \times h}$ is to assign the average pixel esteem $A_{cu_{n}}$ from Eq.~(\ref{eq::linear model}) to the complete area $CU_{n}$ with the size of $\{w_n \times h_n\}$.
Until now, we finish projecting the partitioning information of intra-predicted frame to the blank image,
The blocking effect can be observed in $M_{seg}$, which corresponds to the distribution of frequency information.
It is worth noting that we merge some CUs visually due to the identical $A_{cu_{n}}$ between them, which is not shown in the HEVC partitioning. 
Even though, it matches better with the strategy of actual bit allocation.

\subsubsection{\textbf{Intra-Mode Mappings}}
Intra-prediction is an operation in video coding to eliminate pixel similarity for intra-frames~\cite{sullivan2012overview}.
It executes the predicted mode selection for each PU based on the least distortion principle to reference samples.
33 angular modes for both luma and chroma channel and two non-directional modes (DC and planar) are involved in HEVC/H.265~\cite{sullivan2012overview}, which exceeds the number of modes in AVC/H.264.
Therefore, HEVC/H.265 explicitly provides better compression efficiency on erasing the spatial information redundancies than AVC/H.264.

To generate intra-mode mappings, we first number predicted modes $Pred_i$ from $0$ to $34$ sequentially.
To project modes information into pictures, we then evenly distribute different values in the interval of $[0,238]$ as the regional pixel esteem according to their serial numbers, as Fig.~\ref{Fig::Framework}(b) shown.
The value interval follows pixel value distribution of common pictures.
Differing from segmentation mappings $M_{seg}$, we fix the scale of PUs as the square of 16 with observing a slight effect on performance.
It is noted that pixels within an identical unit share an uniform mode.
At last, we cluster and reshape the comprehensive PUs to the uniform size of the rest of referenceless features,
\begin{equation}
\label{Eq::prediction}
M_{intra} = \O_{r}([A_{pu_1}]_{16 \times 16}, [A_{pu_2}]_{16 \times 16}, ..., [A_{pu_n}]_{16 \times 16}),
\end{equation}
where $n$ is the quantity of PUs in the intra-mode mappings.
The complete extraction of $M_{intra}$ is visualized in Fig.~\ref{Fig::Framework}(b).

\begin{figure*}[htb]
\minipage{1\textwidth}
\includegraphics[width=1\linewidth]{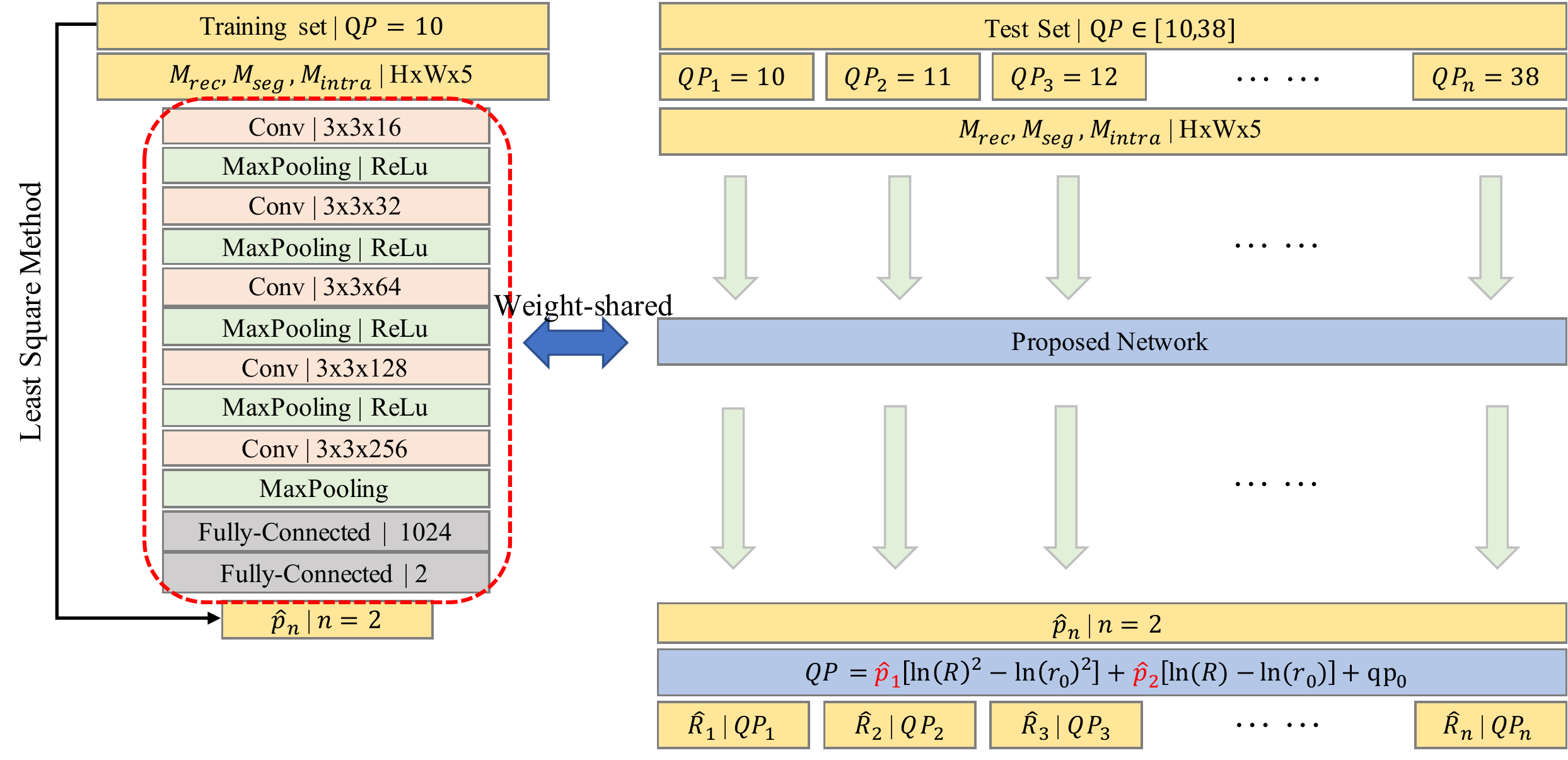}
\endminipage
\caption{Architecture of proposed network and the complete procedures of training and testing.}
\label{fig::network}
\end{figure*}

\section{Architecture of Proposed Network}
\label{Sec::Architecture of Proposed Network}
We propose a convolutional neural network (CNN) to learn the connection between R-QP relationship and hybrid referenceless features, which can be seen as a typical regression problem.
To highlight our main contributions on modeling optimization and features enhancement, we give up the adoption of deeper and more effective neural network but only 5-layers architecture, which is shown in Fig.~~\ref{fig::network}.
We adopt $Adam$ optimizer \cite{kingma2014adam} to train at the learning rate of $0.0001$. 
$ReLu$ as activated function is configured at the output of first four layers.
To normalize the input and output, we use the image normalization and $Standard Scaler$ to process the complete referenceless features and model variables separately. 
Batch size sets up to 10 and the number of epochs is 100.
The loss function is designed as follow,
\begin{equation}
\label{Eq::loss function}
\bm{L}(\Theta) = \frac{1}{n} \sum_{i=1}^n ||\hat{p_n} - p_n||^2.
\end{equation}
where $\hat{p_n}$ and $p_n$ are the predicted model parameters and ground truth ($n$ is number of parameters).
It trains at NVIDIA GeForce GTX 1080 GPU around 22 hours for each task.


\section{Experimental Results}
\label{Sec::Experimental Results}
\subsection{Simulation Setup}
DIV2K dataset \cite{agustsson2017ntire} includes 900 high-resolution images in varying scales.
To fulfill the experiments, we crop 800 images into the desirable patches (512x512 and 768x768) stochastically as training/validation sets and the rest of images as test set.
The ratio of training and validating samples is $90\%$ and $10\%$.
As shown in Fig.~\ref{fig::network}, we train the network parameters $\Theta$ strictly in the single quantized level ($QP=10$), which are applicable for circumstances in the rest of quantized levels ($QP \in [10,38]$) of identical content frames.
As the labels of network, model parameters $p_n$ are calculated by the least square method~\cite{mansard1980measurement}.
To validate the generalization of PmR-QP, we test on two frame resolutions and execute the features extraction at HM 16.9 platform \cite{sullivan2012overview}.

We take the related bitrate estimation error $\delta$ as the measure of accuracy in bitrate estimation as followed,
\begin{equation}
\label{eq::distortion}
  \delta = \frac{R(QP) - \hat{R}(QP)}{R(QP)} \times 100\%.
\end{equation}  

\subsection{Performance of PmR-QP in Bitrate Estimation}
\begin{table}[h]
\setlength{\tabcolsep}{0.8 mm}{
\begin{center}
\caption{Comparison of PmR-QP and the linear modeling solution in bitrate estimation}
\begin{tabular}{ccccccc}
\hline
\multirow{2}{*}{Frame Size} & \multirow{2}{*}{Model} & \multirow{2}{*}{Features} & \multirow{2}{*}{$P_0$}      &  \multicolumn{3}{c}{Estimated Error $\delta$}     \\
\cline{5-7}
              & & & &       30\%  & 20\%  & 10\%                                  \\
\hline
\multirow{3}{*}{512x512} & Linear & $M_{rec}$  & -      & 80.70\%  & 63.14\%  & 35.81\%             \\
& PmR-QP &  All  & \checkmark    & 87.92\%  & 79.11\%  & 60.55\%                         \\
\cline{2-7}
& \multicolumn{3}{c}{\textbf{Improvement}} & \textbf{+7.22\%} & \textbf{+15.97\%} & \textbf{+24.74\%} \\
\hline
\multirow{3}{*}{768x768} & Linear & $M_{rec}$  & -            & 83.56\%  & 67.40\%  & 38.61\%             \\
& PmR-QP &  All  & \checkmark    &  90.43\%  & 81.97\%  & 63.07\%                         \\
\cline{2-7}
& \multicolumn{3}{c}{\textbf{Improvement}} & \textbf{+6.87\%} & \textbf{+14.57\%} & \textbf{+24.46\%} \\ 
\hline
\label{tab:overall_result}
\end{tabular}
\end{center}}
\end{table}

We first evaluate the entire performance improvement of PmR-QP compared with the linear modeling solution. 
Table~\ref{tab:overall_result} shows the accuracy of bitrate estimation by the proposed and linear prediction-based modeling method.
Here, PmR-QP employs the optimized quadratic R-QP modeling function to learn their parameters $\hat{p_n}$ from the overall hybrid referenceless features.
For a fair comparison of modeling function and training features, the linear solution is trained by the proposed network as well.
The results in Table~\ref{tab:overall_result} show that PmR-QP achieves $87.92\%$, $79.11\%$ and $60.55\%$ of samples' estimated error $\delta$ within $30\%$, $20\%$, and $10\%$ in 512x512, respectively.
$14.33\%$, $22.90\%$, and $29.83\%$ precision improves in each error region, which shows that PmR-QP outperforms the linear solution.
In 768x768, the prediction performance of PmR-QP is better than in 512x512 that $90.43\%$, $81.97\%$, and $63.07\%$ in each error region.
$5.65\%$, $13.88\%$, and $23.57\%$ rises are bought by PmR-QP to the linear one.
In general, PmR-QP significantly promotes the accuracy of bitrate estimation by $26\%$ (within $10\%$ error) on average in both resolutions over the linear solution.

\subsection{Ablation Studies of PmR-QP Method}
\label{Sec::Ablation Studies of PmR-QP Method}
\subsubsection{\textbf{Improvements in Quadratic R-QP Modeling Function}}
\begin{table}[h]
\setlength{\tabcolsep}{0.8 mm}{
\begin{center}
\caption{Performance of R-QP modeling function}
\begin{tabular}{ccccccc}
\hline
\multirow{2}{*}{Frame Size} & \multirow{2}{*}{Model} & \multirow{2}{*}{Features} & \multirow{2}{*}{$P_0$}      &  \multicolumn{3}{c}{Estimated Error $\delta$}     \\
\cline{5-7}
              & & & &       30\%  & 20\%  & 10\%                                  \\
\hline
\multirow{3}{*}{512x512} & Linear & \multirow{2}{*}{All}  & \multirow{2}{*}{\checkmark}      & 90.01\%  & 75.19\%  & 43.45\%             \\
& Quadratic &    &     & 87.92\%  & 79.11\%  & 60.55\%                         \\
\cline{2-7}
& \multicolumn{3}{c}{\textbf{Improvement}} &  \textbf{-2.09\%} & \textbf{+3.92\%} & \textbf{+17.1\%} \\
\hline
\multirow{3}{*}{768x768} & Linear & \multirow{2}{*}{All}  & \multirow{2}{*}{\checkmark}             & 91.64\%  & 78.24\%  & 45.38\%             \\
& Quadratic &    &      & 90.43\%  & 81.97\%  & 63.07\%                         \\
\cline{2-7}
& \multicolumn{3}{c}{\textbf{Improvement}} & \textbf{-1.21\%} & \textbf{+3.73\%} & \textbf{+17.69\%} \\ 
\hline
\label{tab:model_result}
\end{tabular}
\end{center}}
\end{table}

To show the superiority of proposed R-QP modeling function, we maintain the identical configuration for the rest optimizations.
The comparison of linear and quadratic modeling function is shown in Table \ref{tab:model_result}, the proportions of samples' estimation error within $30\%$, $20\%$, and $10\%$ are $87.92\%$, $79.11\%$, and $60.55\%$ in 512x512 along with $90.43\%$, $81.97\%$, and $63.07\%$ in 768x768 by utilizing quadratic modeling function.
Compared with the linear function, the increasing precision is $-2.09\%$, $3.92\%$, and $17.10\%$ in 512x512 along with $-1.21\%$, $3.73\%$, and $17.69\%$ in 768x768.
It proves the fidelity of quadratic modeling function at characterizing R-D relationships, especially in preciser estimation scenarios.

\subsubsection{\textbf{Improvements in Modeling Function Simplification}}
\label{subsubsec::Improvements in Modeling Function Simplification}

\begin{table}[h]
\setlength{\tabcolsep}{0.6 mm}{
\begin{center}
\caption{Comparison of simplifying by operational point $P_0$ or not}
\begin{tabular}{ccccccc}
\hline
\multirow{2}{*}{Frame Size} & \multirow{2}{*}{Model} & \multirow{2}{*}{Features} & \multirow{2}{*}{$P_0$}      &  \multicolumn{3}{c}{Estimated Error $\delta$}     \\
\cline{5-7}
              & & & &       30\%  & 20\%  & 10\%                                  \\
\hline
\multirow{3}{*}{512x512} & \multirow{2}{*}{Quadratic} & \multirow{2}{*}{All} & -      & 73.59\%  & 56.21\%  & 30.72\%             \\
&  &    & \checkmark     & 87.92\%  & 79.11\%  & 60.55\%                         \\
\cline{2-7}
& \multicolumn{3}{c}{\textbf{Improvement}} & \textbf{+14.33\%} & \textbf{+22.90\%} & \textbf{+29.83\%} \\
\hline
\multirow{3}{*}{768x768} & \multirow{2}{*}{Quadratic} & \multirow{2}{*}{All} & -             & 84.78\%  & 68.09\%  & 39.50\%             \\
&  &    & \checkmark    & 90.43\%  & 81.97\%  & 63.07\%                         \\
\cline{2-7}
& \multicolumn{3}{c}{\textbf{Improvement}} & \textbf{+5.65\%} & \textbf{+13.88\%} & \textbf{+23.57\%} \\ 
\hline
\label{tab:Performance on the observed operational point}
\end{tabular}
\end{center}}
\end{table}

As known, higher-order models can improve the estimation accuracy of the R-QP relationship, however, blindly improving the number of orders would complicate the learning-based estimated process, even though they can fit complex data well due to the increasing number of model parameters $p_n$.
The notion of model simplification is to inherit an operational point from the one-pass coding, using it to reduce model parameters $p_n$.
Table~\ref{tab:Performance on the observed operational point} shows a comparison of quadratic modeling function based estimated methods with or without model simplification.
In 512x512, the improvement by adopting model simplification is $14.33\%$, $22.90\%$, and $29.83\%$, while the estimated error $\delta$ is below $30\%$, $20\%$, and $10\%$, respectively.
In 768x768, the improvement is $5.65\%$, $13.88\%$, and $23.57\%$ corresponding to each error region.
Model simplification brings the most significant promotion in bitrate estimation, which is $9.99\%$, $18.39\%$, and $26.7\%$ on average in each error region.
It indicates that most deterioration of estimation essentially originates from the capacity of neural network training.

\subsubsection{\textbf{Improvements in the Hybrid referenceless features}}
%
\begin{table}[h]
\setlength{\tabcolsep}{0.8 mm}{
\begin{center}
\caption{Comparison of different combinations of hybrid referenceless features}
{
\begin{tabular}{cccccc}
\hline
\multirow{2}{*}{Frame Size} & \multirow{2}{*}{Model} & \multirow{2}{*}{Patterns}         &  \multicolumn{3}{c}{Bitrate Estimated Error $\delta$}     \\
                  &  &   & 30\%  & 20\%  & 10\%                                  \\
\hline
\multirow{7}{*}{512x512} & \multirow{7}{*}{Quadratic} & $M_{seg}$           & 82.24\%  & 71.07\%  & 51.17\%             \\
&& $M_{intra}$        & 80.88\%  & 70.22\%  & 51.01\%                         \\
&& $\{M_{seg},M_{intra}\}$        & 82.80\%  & 72.78\%  & 52.18\%                         \\
&& $M_{rec}$        & 84.96\%  & 76.41\%  & 56.96\%                         \\
&& $\{M_{rec},M_{seg}\}$        & 86.54\%  & 77.32\%  & 58.03\%                         \\
&& $\{M_{rec},M_{intra}\}$        &  86.53\% & 77.53\%  &  58.43\%                        \\
&& All        & \textbf{87.61\%} & \textbf{79.11\%} & \textbf{60.55\%}                         \\
\hline
\multirow{7}{*}{768x768} & \multirow{7}{*}{Quadratic} & $M_{seg}$           & 81.92\%  & 71.88\%  &  52.64\%            \\
&& $M_{intra}$        &  79.89\% & 69/88\% & 51.81\%                        \\
&& $\{M_{seg},M_{intra}\}$        & 84.13\%  & 73.80\%  & 53.43\%                         \\
&& $M_{rec}$        & 84.48\%  & 75.18\%  & 56.19\%                         \\
&& $\{M_{rec},M_{seg}\}$        & 87.16\%  & 78.53\%  & 60.32\%                         \\
&& $\{M_{rec},M_{intra}\}$        &  88.82\% & 79.37\%  &  60.42\%                        \\
&& All       & \textbf{90.43\%} & \textbf{81.97\%} & \textbf{63.07\%}                         \\
\hline
\label{tab:bitstream_error}
\end{tabular}
}
\end{center}}
\end{table}
This section discusses the performance of hybrid referenceless features.
In prior researches, pixel domain features or bitstream features are adopted more often due to the inconsistency between the two different domains.
We find a method to extract and unify multi-domains coding features and observe their R-QP estimation performance in Table~\ref{tab:bitstream_error}, which demonstrates the superiority of entire features combination over other ways to combine features.
It indicates the existence of non-overlapping information from different domains to achieve the performance-boosting of estimation.
Compared with texture domain features $M_{rec}$, the entire features combination can bring out $2.98\%$ and $6.54\%$ improvements on average in each resolution.
Meanwhile, it also shows a better improvement compared with other coding bitstream domains, e.g., $M_{seg}$ or $M_{intra}$.
Note that our method at feature extraction always performs better in higher resolution scenarios.

Besides, we dig out more facts from the comparison of different domain features.
In single domain level, $M_{rec}$ outperforms $M_{seg}$ or $M_{intra}$ in estimation, which reveals a fact that richer information is provided by texture of frames.
Even though, $M_{seg}$ and $M_{intra}$ show a very close outcome due to partial over-lapping texture information in these coding domains.
Moreover, $M_{seg}$ replenishes the knowledge of partitioning structure and $M_{intra}$ describes the similarity of neighboring blocks.
They represent redundancy in different aspects.
By combining them with $M_{rec}$ one by one, we can observe obvious growths and the best performance while combining all.
It is noted that only $M_{seg}$ as input can achieve $51.1\%$ and $52.64\%$ (within $10\%$ error) in each resolution, which adopts the lowest network and coding complexity.

\subsection{Observation on Different Behaviors in Quantization}
We investigate R-QP curves by adopting different aforementioned optimized methods to further claim the outperformance of PmR-QP.
We chose two typical data samples in different quantifying changing circumstances.
It is observed that the best fitting actual R-QP curve is given by PmR-QP employed the entire hybrid referenceless features and simplified R-QP modeling function compared with other methods, especially significantly outperform the conventional method.
The worst performance is provided by PmR-QP without simplifying R-QP function, which expresses the deterioration of results originated from the training process.
However, due to fitting limits of linear modeling function, model simplification cannot bring any obvious gains.
The performance of the linear modeling based solution is as good as the quadratic method in uniform quantifying changes but much worse in non-uniform circumstances.

\section{Conclusion}
\label{Sec::Conclusion}
In this paper, we propose a referenceless PmR-QP to predict bitrate in given frames' quality information precisely throughout only one-pass coding.
This method is built on the root of the global rate-control paradigm, which takes advantage of inborn systemic superiority over block-level paradigm in Internet-based multimedia services.
Moreover, it efficiently tackles the defects of prior global rate-control methods.
In detail, first, we derive the quadratic R-QP modeling function from Cauchy-based distribution on the entropy of DCT's coefficients, which is better at fitting the relationship between rate and level of quantization than linear function and applicable in most real cases.
Second, efficiently utilizing the coding information, we involve an operational point to simplify the proposed modeling function.
Third, PmR-QP significantly enhances the description of characteristics between R-QP and the frames' content information by exploring and unifying bitstream features from multiple coding domains.
Extensive experiments and ablation studies demonstrate the global improvements in PmR-QP, and the performance-boosting from each optimized step.
Generally, PmR-QP can achieve $24.60\%$ decreases on samples' bitrate estimating error lower than $10\%$ on average compared with the state-of-the-art.

In the future, we intend to expand our work to the inter-prediction level by exploring features of motion estimation.
Likely, the similarity of successive inter-frames can be represented by the accumulated motion estimation features since they have been applied in other compressed video tasks.
It is possible to aggregate these features as elements of hybrid referenceless features to learn R-QP information of intra- and inter-frames.

\bibliographystyle{unsrt}
\bibliography{references.bib}

\end{document}